\documentclass[runningheads]{llncs}
\usepackage[T1]{fontenc}

\renewcommand{\th}{$^{\mbox{\scriptsize{\textrm{th}}}}$}

\usepackage{graphics}
\usepackage{url}
\usepackage{algorithm}
\usepackage[noend]{algpseudocode}

\begin{document}
	\title{Sorted Range Selection and Range Minima Queries}
	
	\author{Waseem Akram\inst{1} \and
		Sanjeev Saxena\inst{1}}
	
	\authorrunning{Akram and Saxena}
	
	\institute{ Dept. of Computer Science and Engineering, 
		Indian Institute of Technology, Kanpur, INDIA-208 016\\
		\email{\{akram,ssax\}@iitk.ac.in}}
	\maketitle           
	\begin{abstract}
		Given an array $A[1: n]$ of $n$ elements drawn from an ordered set, the sorted range selection problem is to build a data structure that can be used to answer the following type of queries efficiently: Given a pair of indices $i, j$ $ (1\le i\le j \le n)$, and a positive integer $k$, report the $k$ smallest elements from the sub-array $A[i: j]$ in order.
		Brodal et al. (Brodal, G.~S., Fagerberg, R., Greve, M., and L{\'o}pez-Ortiz, A., Online sorted range reporting. Algorithms and Computation (2009) pp.~173--182)
		introduced the problem and gave an optimal solution. After $O(n\log n)$ time for  preprocessing, the query time is $O(k)$. The space used is $O(n)$.
		
		In this paper, we propose the only other possible optimal trade-off for the problem. We present a linear space solution to the problem that takes $O(k\log k)$ time to answer a range selection query. The preprocessing time is $O(n)$. Moreover, the proposed algorithm reports the output elements one by one in non-decreasing order.
		Our solution is simple and practical.

		We also describe an extremely simple method for range minima queries
(most of whose parts are known) which takes al most (but not exactly)
linear time. We believe that this method may be, in practice, faster
and easier to implement in most cases.
		
		\keywords{Range Minimum Query \and Range Reporting \and Algorithms \and Data Structures}
	\end{abstract}
	
	\section{Introduction}
	The range minimum query (RMQ) problem is a well-studied problem \cite{FISCHER11,FISCHER09,AMIR92,SADAKANE07,SAXENA09,GABOW84}. Given an array, the RMQ problem is to find the position (i.e., index) of the smallest element (in an index range).
	In this paper, we study the sorted range selection problem \cite{BRODAL09}, a
	generalization of the range minimum query problem.
	Given an input array $A[1:n]$ of $n$ elements drawn from an ordered set, the problem is to preprocess the array $A[1:n]$ so that the queries of the following type can be answered efficiently \cite{BRODAL09}: 
	\begin{quote}
		Given a pair of indices $i,j$ with $(1\le i\le j\le n)$ and a positive integer $k$, report the $k$ smallest elements in the index range $[i,j]$ in sorted order.
	\end{quote}
	Brodal, Fagerberg, Greve, and L{\'o}pez-Ortiz \cite{BRODAL09} introduced the problem and gave a linear space data structure with $O(k)$ query time. The preprocessing time to build the structure is $O(n\log n)$. 
	By reporting all $n$ elements ($k=n$), one can sort the elements in $O(n)$ time; thus, $\Omega(n\log n)$ preprocessing time is required, and their algorithm is optimal. 
	The solution uses a fairly complicated result due to Frederickson and Johnson \cite{FREDERICKSON82}.
	
	We propose the only other possible optimal trade-off for the sorted range selection problem. The preprocessing time is $O(n)$ with linear space, and the query time is $O(k \log k)$. Our solution is also optimal as by reporting all elements in the array ($k=n$), we can sort the elements of the input array in $O(n \log n)$ time.
	
	Note that our solution offers the only other possible optimal trade-off for the problem. If the preprocessing time is $p(n)$ and the query algorithm takes $q(n)$ time to report any element in the query range in the worst case, then we can sort $n$ elements in $p(n) + \sum_{i=1}^{n} q(n)$ time. Thus, if $p(n)=o(n\log n)$, then $q(n)=\Omega(\log n)$.
	Our algorithm is extremely simple and only uses range minimum queries in addition to the usual binary heap. Moreover, the algorithm reports the output elements one by one in
	non-decreasing order.

	A related and more general problem that has been studied in the past is the range selection problem \cite{BRODAL09,AFSHANI11}, where the output
	elements are not required to be reported in sorted order. Brodal et
	al. \cite{BRODAL09} suggested that an array can be preprocessed using
	linear space and time to answer a range selection query in $O(k)$ time
	in the RAM model. Brodal et al. \cite{BRODAL09} also suggested a method to solve the problem in the pointer machine model by using the priority search tree \cite{CREIGHT85} and Frederickson's $O(k)$-time algorithm \cite{FREDERICKSON93} for finding
	the $k$ smallest elements in a binary heap. The resultant structure takes linear space and can report $k$ smallest elements in $O(\log n + k)$ time.
	Skala \cite{Skala13} presented a survey of array range query problems.

	\section{Proposed Solution}
	We use the following notations.
	Let $A[1:n]$ be the input array of $n$ elements drawn from a totally ordered set. 
	For any $i, j\in \{1,2,...,n\}$ with $i\le j$, $A[i:j]$ denotes the sub-array starting at index $i$ and ending at index $j$, and $A[i]$ denotes the array element at index $i$. For any two parameters $i$ and $j$, $1\le i\le j\le n$, the range minimum query, denoted by $RMQ(i,j)$, is to find the index of a minimum element among $A[i], A[i+1],... A[j]$.
	
	Our solution is based on the following observation.
	\begin{quote}
		Consider any two fixed indices $i\mbox{ and }j$ with $i\le j$. If
		$A[r]$ is the smallest element in the sub-array $A[i: j]$, then the
		next smallest element in $A[i:j]$ will either be in the sub-array
		$A[i:r-1]$ or in the sub-array $A[r+1: j]$.
	\end{quote}
	We first preprocess the given array $A[1:n]$ for the range minimum queries (RMQ) \cite{GABOW84,SCHIEBER88}. The preprocessing takes $O(n)$ time and space. For each subsequent range minimum query, $RMQ(i,j)$ with $i\le j$, the RMQ data structure returns the index $r$ of the minimum element $A[r]$ in the subarray $A[i:j]$ in $O(1)$ worst-case time.

In Section~3, we describe, for completeness, an extremely simple
method for range minima queries which takes $O(n \log^{k} n)$
preprocessing time, for any $k$\footnote{let $\log^{(1)}n=\log n,\log^{(k+1)}n=\log\left(\log^{(k)}n\right)$} and can answer queries in
$O(1)$ time, as $\log^{(5)}n\leq 2$ for $n\leq 2^{65536} \approx 10^{19728}$, this method takes almost (but not exactly) linear time. 
We believe that the method will be, in practice, faster and easier to implement in most cases.

Assume we are to report the $k$ smallest elements from the sub-array $A[i:j]$.
	An RMQ will give the index $r$ of the smallest element in the subarray $A[i:j]$ in $O(1)$ time. We report $A[r]$ as the smallest element in $A[i:j]$. We then split the interval into two pieces $[i, r-1]$ and $[r+1, j]$ and use RMQs to find the minimum element in each instance. We insert these elements into (an initially empty) binary heap (see, e.g., Chapter $6$ in \cite{Cormen09}). 
	We keep performing the following step until $k$ elements are reported or the min heap becomes empty:
	\begin{quote}
		Remove the minimum element from the heap and report it as the next smallest element in the subarray $A[i:j]$. Split the interval of the minimum element into two pieces and insert their minimum elements into the heap.
	\end{quote}
	Algorithm $1$ is the pseudo-code of the sorted range selection query procedure.
	\begin{algorithm}
		\caption{Sorted Range Selection}\label{euclid}
		\hspace*{\algorithmicindent} \textbf{Input} $i,j,k$ with $1\le i,j \le n$ and $1\le k \le j-i+1$\\
		\hspace*{\algorithmicindent} \textbf{Output} $k$ smallest elements in the subarray $A[i:j]$
		\begin{algorithmic}[1]
			\State $Q\gets \phi$ \Comment{Initializing min-heap}
			\State $x_l, x_r \gets -1$ \Comment{index variables}
			\State $r\gets RMQ(i,j)$ 
			\State insert the element $A[r]$ into the heap $Q$. \Comment{$[i,j]$ is stored as satellite data}
			\Repeat
			\State delete the minimum element $A[x]$ from heap $Q$
			\State report $A[x]$ as $t^{th}$ smallest element if the current iteration is $t^{th}$ iteration
			\State let $[p,q]$ be the satellite information associated with $A[x]$
			\State if $p<x$, then $x_l\gets RMQ(p, x-1)$ 
			\State if $x<q$, then $x_r \gets RMQ(x+1, q)$
			\State insert $A[x_l]$ (resp. $A[x_r]$) into the heap $Q$ if $x_l \ne -1$ (resp. $x_r \ne -1$).
			\State $x_l, x_r \gets -1$ \Comment{initializing for next iteration}
			\Until{$k$ elements have been reported or the heap $Q$ gets exhausted}
		\end{algorithmic}
	\end{algorithm}
	We illustrate Algorithm $1$ with an example in Subsection $2.1$.
	\begin{remark}
		As the next smallest element of $A[p:q]$ will be either in
		$A[p:x-1]$ or $A[x+1:q]$, the next smallest element (to be reported)
		will always be in a heap.
	\end{remark}	
	\begin{remark}
		Our algorithm reports or outputs the required elements one by one in non-decreasing order. 
	\end{remark}
	As in each iteration, we are deleting one element from the heap and
	inserting at most two more; after $i$ iterations, we will have at
	most $i+1$ elements in the heap. Insertion or deletion in a binary heap takes $O(\log s)$ time, where $s$ is the size of the heap before the operation. Therefore, each insertion or deletion of an element in $i$\th iteration will take $O(\log i)$ time, 
	or the total time will be O($\sum_{i=1}^k \log i)=O(k \log k)$. We have the following theorem.
	\begin{theorem}
		An array $A[1:n]$ of $n$ elements drawn from a totally ordered set can
		be preprocessed so that given a pair of indices $i,j$ with $1 \le i\le
		j\le n$ and a parameter $k$, we can report the $k$ smallest elements
		in the subarray $A[i:j]$ in $O(k\log k)$ time. The preprocessing takes
		$O(n)$ space and time. 
	\end{theorem}
	\qed
	\begin{remark}
		If elements in the input array $A[1:n]$ are integers from a set
		$[1..U]$, then we can use van Emde Boas structure (see Chapter $20$ in
		\cite{Cormen09}) for implementing the pool. As a result, the query
		time will become $O(k\log\log U)$.
	\end{remark}

\section{Range Minimum Query}
In range minima query we are given an array $A[1:n]$ which we can
preprocess. We have to answer queries of kind:
\begin{quote}
RMQ$(i,j)$: Find the index of the smallest element in
$A[i],A[i+1],{\ldots} ,A[j-1],A[j]$
\end{quote}
Query time should be $O(1)$.

Assume that after preprocessing, for each position $i$ in the array we
know the (index of) minimum element in each of the following cases
$$A[i:i+1],A[i:i+2],A[i:i+2^2],{\ldots} ,A[i:i+2^j]$$
for $i+2^j\le n$.

Then query RMQ$(i,j)$ to find the minimum element in $A[i:j]$ can be
answered as follows
\begin{enumerate}
\item Let $r$ be the largest integer s.t., $i+2^r\le j$, or
equivalently, $2^r\leq j-i$

		\begin{remark}
			$r$ is the index of the most significant bit in binary
representation of $j-i$.
		\end{remark}

\item Using precomputations, we can find the (index of) minimum element
in $A[i:i+2^r]$. 

\item If $j'=j-2^r$, then again using precomputations, we can find the
(index of ) minimum element in $A[j':j'+2^r]$ or $A[j':j]$.

\item The (index of) the required minimum element is (the index of)
smaller of these two values, and hence can be found in $O(1)$ time.
\end{enumerate}

As $r$ is the largest integer s.t., $i+2^r\le j$, or $i+2^{r+1}>j$, or
(subtracting $2^r$), $i+2^r>j-2^r=j'$. Thus, the union of intervals
$[i:i+2^r]$ and $[j':j]$ is $[i:j]$ (portion $[j':i+2^r]$ is common to
both intervals). Thus, we are computing minimum of elements exactly in
the range $i{\ldots} j$ (some elements are however considered twice).

Let us now look at the precomputation. Assume for each $1\le i\le n$
(and for some $j$) we have computed the (index of) minimum element in
$A[i:\min\{i+2^j,n\}]$. Then we can compute the (index of) minimum element in
$A[i:\min\{i+2^{j+1}],n\}$, for each $1\le i\le n$ as follows:
\begin{enumerate}
\item If $i+2^j\geq n$, then $i+2^{j+1}\geq n$. Or (index of) minimum
element in $A[i:\min\{i+2^j,n\}]$ is also the (index of) minimum element in
$A[i:\min\{i+2^{j+1}],n\}$.

\item Else ($i+2^j<n$) let $i'=i+2^j$. By hypothesis (precomputation)
we know the (index of) minimum element in $A[i':\min\{i'+2^j,n\}]$ (or
$A[i+2^j:\min\{i+2^{j+1},n\}]$).

\item The (index of) minimum element in $A[i:\min\{i+2^{j+1},n\}]$ is the
(index of) the smaller of the two numbers:
\begin{quote}
minimum element in $A[i:i+2^j]$ and the minimum element in
$A[i+2^j:\min\{i+2^{j+1},n\}]$.
\end{quote}
\end{enumerate}
Thus, for each $1\leq i\leq n$, we can find $A[i:\min\{i+2^{j+1}],n\}$
in $O(1)$ time, or for all $1\leq i\leq n$ in $O(n)$ time.
As $0\leq j\leq \log n$, entire precomputation takes $O(n\log n)$ time.

The complete algorithm is:
\begin{algorithm}
	\caption{}\label{euclid}
	\begin{algorithmic}[1]
		\For{$i\gets1$ to $n$}
			\State $B_1[i]\gets i$;
			\If{$i+1\leq n$ and
				$(A[i+1]<A[i])$}
				\State $B_1[i]\gets i+1$;
			\EndIf
		\EndFor
		\For{$k \gets 2$ to $\log n$}
			\For{$i\gets1$ to $n$}
				\State $B_k[i] \gets B_{k-1}[i]$;
				\If{$i+2^k< n$}
				\State $r \gets B_{k-1}[i]$ and $s \gets B_{k-1}[i+2^k]$;
				\If{$(A[s]<A[r])$}
					\State $B_k[i] \gets s$;
				\EndIf
				\EndIf
			\EndFor
		\EndFor
	\end{algorithmic}
\end{algorithm}
~\\
~\\
~\\
~\\
\newpage

Thus we have \cite{Bender00,Bender05,Fischer06,rmqwiki}

\begin{lemma} \label{wiki:lemma}
An array $A[1:n]$ can be preprocessed in $O(n\log n)$ time and space
such that queries of kind:
\begin{quote}
RMQ$(i,j)$: Find the index of the smallest element in
$A[i],A[i+1],{\ldots} ,A[j-1],A[j]$
\end{quote}
can be answered in $O(1)$ time.
\end{lemma}

\subsection{Linear space solution}

The space can be reduced to $O(n)$ as follows \cite{rmqwiki}.
\begin{enumerate}
\item The array $A[1:n]$ is conceptually split into $n/\log n$ blocks of
size $\log n$. 

\item The minimum of each block of $\log n$ elements is computed, 
in $O(\log n)$ time. As there are $n/\log n$ blocks, total time is
$O(n)$ overall. We also compute the prefix minimum (smallest element from
start of block) and suffix minimum (smallest element till end of the
block). This can also be done in same time bounds.

\item These minima are stored in another array of length $n/\log n$, say
$S\left[1:\frac{n}{\log n}\right]$. 

\item The array $S$ is preprocessed as per Lemma~\ref{wiki:lemma}.

As $S$ has $n/\log n$ elements, it will take $O\left( (n/\log n)
\log\left(n/\log n)\right)\right)=O(n)$ time and space.
\end{enumerate}
Thus, preprocessing time and space is $O(n)$.

A query $RMQ(l,r)$ when two elements are not in the same block can be
answered in $O(\log n)$ time as follows:
\begin{enumerate}
\item Find $i=\lfloor l/\log n\rfloor$ and $j=\lfloor r/\log
n\rfloor$, the the block(s) containing the two indices.

\item If $i<j$, then find $k=RMQ(i+1,j-1)$. 

Basically, the minima of all blocks contained completely inside the
range is computed using a query to the data structure built over array
$S$ in $O(1)$ time.

\item If $i\neq j$, then as we know the suffix minima at location $l$
in block $i$ and prefix minima at location $r$ in block $j$. Comparing
these two elements with the element computed in previous step, we get the
overall minimum.
\end{enumerate}

We are left with the case, when both $l$ and $r$ are in the same
group, i.e., when $i=j$. This is the usual range minima query,
restricted to a block. 

If we preprocess each block (independently and separately) for range
minima query using algorithm of Lemma~\ref{wiki:lemma}, the time for
each block is $O((\log n) \log ( \log n))=O(\log  n \log\log n)$. As
there are $n/\log n$ blocks, total time is $O((n/\log n)(\log
n\log\log n))=O(n\log\log n)$. 

Thus we have:
\begin{corollary} \label{wiki:cor}
An array $A[1:n]$ of $n$ elements drawn from a totally ordered set can
be preprocessed in $O(n\log\log n)$ time and $O(n)$ space such that
range minima queries can be answered in $O(1)$ time.
\end{corollary}

If we use the method of Cor~\ref{wiki:cor} for preprocessing each
block for range minima queries, the preprocessing time 

Using $k$-level structure, the preprocessing time can be easily
reduced to $O(n\log^{(k)} n)$. We next describe an almost linear time
solution. We have to only consider the case when the two elements are in
same block. The complete preprocessing algorithm is:

\subsection{Almost Linear Time Method}

\begin{enumerate}
\item The array $A[1:n]$ is conceptually split into $n/\log n$ blocks of
size $\log n$ (as before).

\item The minimum of each block of $\log n$ elements is computed, 
in $O(\log n)$ time. As there are $n/\log n$ blocks, total time is
$O(n)$ overall.
\begin{enumerate}
\item We also compute the prefix minimum (smallest element from
start of block) and suffix minimum (smallest element till end of the
block). This can also be done in same time bounds.
\item If number of elements in a block is less than $32$ (a constant), we
preprocess each block for range minima queries using the method 

\end{enumerate}
\item These minima are stored in another array of length $n/\log n$, say
$S\left[1:\frac{n}{\log n}\right]$. 

\item The array $S$ is preprocessed as per Lemma~\ref{wiki:lemma}.

As $S$ has $n/\log n$ elements, it will take $O\left( (n/\log n)
\log\left(n/\log n)\right)\right)=O(n)$ time and space.
\end{enumerate}
Thus, preprocessing time and space is $O(n)$.

A query $RMQ(l,r)$ when two elements are not in the same block can be
answered in $O(\log n)$ time as follows:
\begin{enumerate}
\item Find $i=\lfloor l/\log n\rfloor$ and $j=\lfloor r/\log
n\rfloor$, the the block(s) containing the two indices.

\item If $i<j$, then find $k=RMQ(i+1,j-1)$. 

Basically, the minima of all blocks contained completely inside the
range is computed using a query to the data structure built over array
$S$ in $O(1)$ time.

\item If $i\neq j$, then as we know the suffix minima at location $l$
in block $i$ and prefix minima at location $r$ in block $j$. Comparing
these two elements with the element computed in previous step, we get the
overall minimum.
\end{enumerate}

It is further possible to reduce the preprocessing time to $O(n)$, but
for that a popular way of doing uses Cartesian tree, Euler tour
traversal, and table look-up
\cite{Fischer06,Schieber88,Berkman93,Berkman1993,Berkman95,Berkman98,Bender00,Bender05}.
	
	\section{Conclusion}
	In this paper, we studied the range selection problem and gave a linear space solution with $O(k\log k)$ query time and $O(n)$ preprocessing time. The output elements are reported individually in non-decreasing order.
	The proposed solution offers the only possible trade-off other than the one given by Brodal et al. \cite{BRODAL09}. Our solution is simple and easy to implement. The data structure of the solution consists of an RMQ structure and a usual binary min heap.
	
	To the best of our knowledge, the sorted range selection problem has not been studied in a dynamic setting. One can consider the problem in the dynamic setting where an update operation can change the element value stored at an index without changing the element values stored at any other indices. We leave this as an open problem.

\bibliographystyle{acm}

\end{document}